\documentclass[3p]{elsarticle}

\usepackage{a4wide}
\usepackage{float}
\usepackage{amsmath}
\usepackage{amssymb}
\usepackage{graphicx}
\usepackage[latin1]{inputenc} 
\usepackage{rotating}
\usepackage{setspace}
\usepackage{subfigure}
\usepackage{sidecap}
\usepackage{upgreek}
\usepackage{microtype}
\usepackage{url}
\usepackage[pdftitle={Time dependence of charge losses at the Si-SiO2 interface in p+n-silicon strip sensors}, pdfauthor={Thomas Poehlsen, Eckhart Fretwurst, Robert Klanner, Joern Schwandt and Jiaguo Zhang}]{hyperref}
\usepackage[all]{hypcap}

\hypersetup{pdflang=en, hidelinks}

\begin{document}
\begin{frontmatter}


\title{Time dependence of charge losses at the Si-SiO$_2$ interface in $p^+n$-silicon strip sensors}


\renewcommand{\thefootnote}{\fnsymbol{footnote}}

\author[]{Thomas Poehlsen \corref{cor1}}
\author[]{Eckhart Fretwurst}
\author[]{Robert Klanner}
\author[]{Joern~Schwandt}
\author[]{and Jiaguo~Zhang}
\cortext[cor1]{Corresponding author. Email address: thomas.poehlsen@desy.de. Telephone: +49 40 8998 4725.}

\address{Institute for Experimental Physics, University of Hamburg, Luruper Chaussee 149, 22761 Hamburg, Germany}

\begin{abstract}

 The collection of charge carriers generated in $p^+n$-strip sensors close to the Si-SiO$_2$ interface before and after 1 MGy of X-ray irradiation has been investigated using the transient current technique with sub-nanosecond focused light pulses of 660~nm wavelength, which has an absorption length of 3.5~$\upmu$m in silicon at room temperature.
 The paper describes the measurement and analysis techniques used to determine the number of electrons and holes collected. Depending on biasing history, humidity and irradiation, incomplete collection of either electrons or holes is observed. The charge losses change with time. The time constants are different for electrons and holes and increase by two orders of magnitude when reducing the relative humidity from about 80~\% to less than 1~\%. An attempt to interpret these results is presented.

\end{abstract}

 \begin{keyword}
  silicon sensors \sep charge losses \sep X-ray-radiation damage \sep humidity \sep biasing history 
\end{keyword}

\end{frontmatter}

\section{Introduction}

 In this paper the time dependence of charge losses close to the Si-SiO$_2$ interface in $p^+n$-silicon strip sensors after changing the bias voltage in environments of different humidities is investigated. Measurements were performed on a non-irradiated sensor and on a sensor irradiated by $\sim$~12~keV X-rays to a dose of 1~MGy. The work is part of the study of X-ray-radiation damage of segmented silicon sensors for the AGIPD detector~\cite{AGIPD} at XFEL, the European X-ray Free-Electron Laser~\cite{XFEL}, where X-ray doses of up to 1~GGy are expected.

 In a previous publication~\cite{Poehlsen:2012} it has been shown that time resolved measurements (TCT - Transient Current Technique) for electron-hole pairs generated by a focussed laser close to the Si-SiO$_2$ interface, allow investigating the electric fields and the properties of possible accumulation layers. For a non-irradiated sensor it has been observed that, after biasing the sensor, the Si-SiO$_2$-interface region is not in steady-state conditions and that the time to reach steady-state conditions depends on humidity: It is about two orders of magnitude shorter for high compared to low relative humidity. In this paper we extend these measurements to a sensor irradiated to an X-ray dose of 1~MGy in order to investigate, how the time development of the electric field in the region close to the sensor surface depends on X-ray-radiation damage.

 The cause of the time and humidity dependence for the non-irradiated sensor was explained by the time dependence of the electric boundary conditions on the surface of the sensor: Changing the bias voltage results in surface fields which cause a redistribution of surface charges until a uniform potential is reached. The strong dependence of the surface resistance on humidity~\cite{Grove:1967} is responsible for the big difference in time constants.

 For given densities of oxide charges and charged interface traps, the electric boundary conditions on the sensor surface   strongly influence the formation of accumulation layers and of the electrical fields close to the Si-SiO$_2$ interface. As the surface boundary conditions can influence the breakdown voltage~\cite{Longoni:1990,Hartjes:2005}, their knowledge is highly relevant for the simulation of segmented silicon sensors~\cite{Richter:1996}, in particular for high X-ray doses and operation in vacuum.

\section{Investigated sensor and measurement technique}

\begin{figure}
	\centering
	\includegraphics[width=8cm]{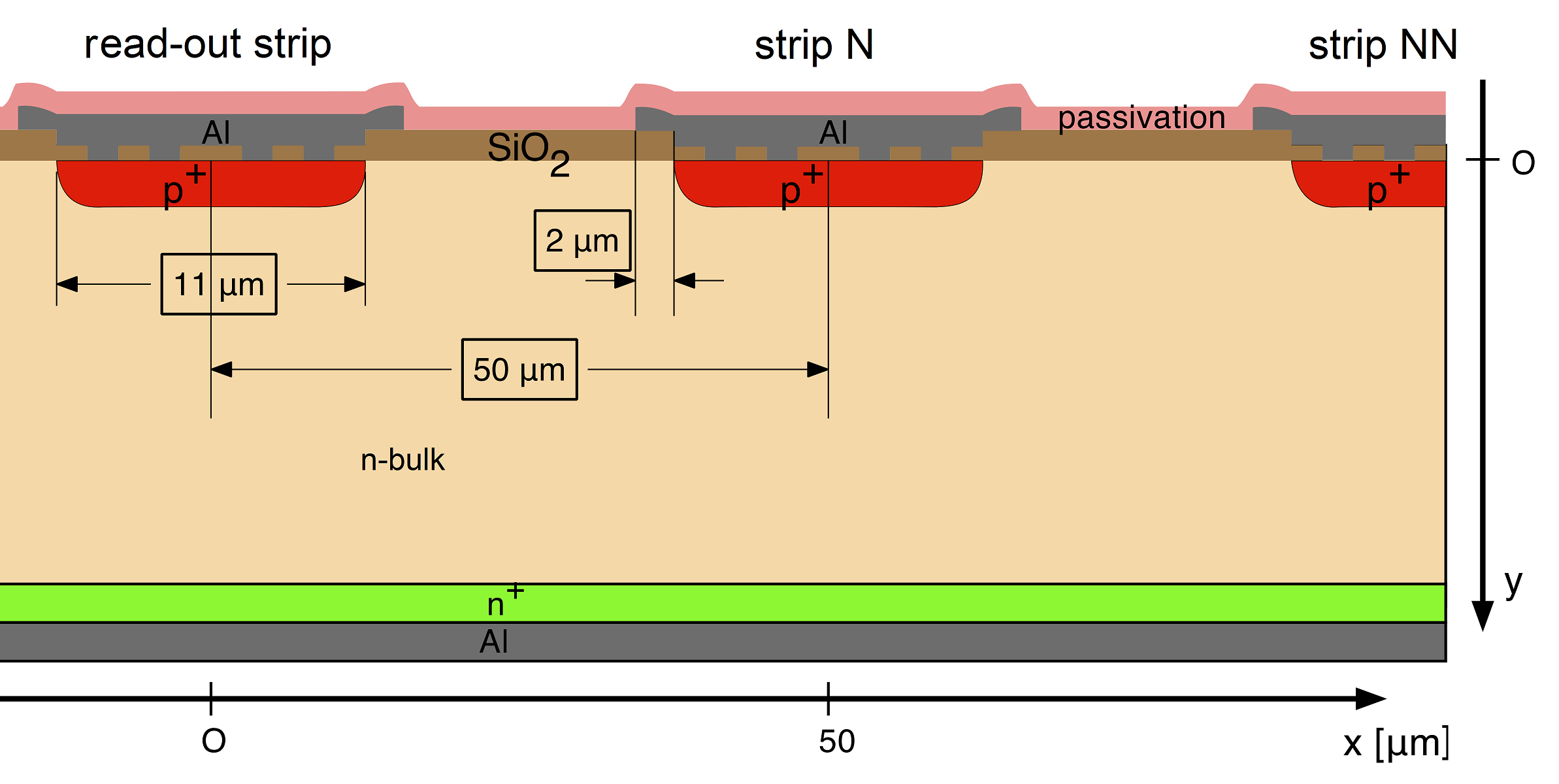}	
       \caption{Schematic layout of the strip region of the DC-coupled Hamamatsu $p^+n$ sensor, and coordinate definition. The drawing is not to scale.}
	\label{fig:sensor}
\end{figure}

 A DC coupled $p^+n$ strip sensor produced by Hamamatsu Photonics~\cite{Hamamatsu} was investigated. Previous studies of the sensor are reported in~\cite{Poehlsen:2012,Becker:Thesis}. The cross section of the sensor is shown in Figure~\ref{fig:sensor} and relevant sensor parameters are listed in Table~\ref{tab:sensors}. Measurements were performed before and after irradiation to a dose of 1~MGy~(SiO$_2$) by $\sim$~12~keV X-rays.
 All measurements were taken at 200~V bias voltage, well above the depletion voltage of 155~V.

\begin{table}[b]
\centering
\begin{tabular}[b]{|c|c|}
\hline
producer 					& Hamamatsu 	\\ \hline
coupling 					& DC 			\\ \hline
pitch 						& 50 $\upmu$m 	\\ \hline
depletion voltage		    & $\sim$~155 V		\\ \hline
doping concentration		& $\sim$~10$^{12}$ cm$^{-3}$	\\ \hline
single strip capacitance 		& 1.4 pF		\\ \hline
gap between $p^+$\,implants	& 39 $\upmu$m 	\\ \hline
width $p^+$\,implant window	& 11 $\upmu$m	\\ \hline
depth $p^+$\,implant  		& unknown 		\\ \hline
aluminium overhang 			& 2 $\upmu$m	\\ \hline
number of strips 			& 128        	\\ \hline
strip length 				& 7.956 mm 		\\ \hline
sensor thickness			& 450 $\upmu$m	\\ \hline
thickness SiO$_2$			& 700 nm     	\\ \hline
passivation layer			& unknown    	\\ \hline
crystal orientation 		& $\langle 1 1 1 \rangle$ \\ \hline
 \end{tabular}
   \caption{Parameters of the Hamamatsu sensor.}
   \label{tab:sensors}
\end{table}

 Pulsed focussed red laser light with a wavelength of 660~nm was used to generate electron-hole ($eh$) pairs in the sensor at the $p^+$-strip side, just below the SiO$_2$ separating the read-out strips. A laser of this wavelength has a penetration depth in silicon of about 3.5~$\upmu$m at room temperature. Hence, the charges were generated close to the Si-SiO$_2$ interface. The light was focussed to a Gaussian with $\sigma \sim 3$~$\upmu$m. The pulses had a duration of 100~ps at FWHM, the number of generated $eh$ pairs was about 100, and the repetition rate 1~kHz.

 The  current signals induced in the read-out strip were read out by DC-coupled Femto HSA-X-2-40 current amplifiers ($\sim$~50~$\Omega$ input impedance, 180~ps rise time) connected to a Tektronix digital oscilloscope with $2.5$ GHz bandwidth (DPO 7254). Neighbouring strips were grounded via 50~$\Omega$ resistors. The  charge $Q_i$ induced on read-out strip $i$ was calculated off-line by integrating the current signal over 16~ns.

 The humidity during the measurements was varied between $\le 1$~\% and $\sim$~85~\%, and the temperature was $\sim$~24$^\circ$C. Humidity and temperature were logged with a TFD128 temperature and humidity data logger. Some earlier measurements were done without data logger, and only approximate values of the humidity are known.

\section{Determination of charge losses}

 This chapter presents the method used to determine the charge losses and shows evidence that in the $p^+n$-strip sensor investigated, depending on the biasing history, humidity and X-ray dose, situations without charge losses, with electron losses and with hole losses close to the Si-SiO$_2$ interface have been observed.

 The electrons and holes generated by the laser drift in the electric field in the sensor thereby inducing currents in the electrodes. The current induced in electrode $i$ by a charge $q$ moving at  position $\vec x$ with  drift velocity $\vec v_{dr}$  can be calculated using the weighting potential $\phi_w^i$:

 \begin{equation}
   I_i = q \cdot \vec E_w^i(\vec x) \cdot \vec v_{dr}(\vec x), \quad \vec E_w^i(\vec x) = \vec \nabla \phi_w^i(\vec x).
\end{equation}

 The weighting potential $\phi_w^i(\vec x)$ is a measure of the capacitive coupling of a unit charge at position $\vec x$ to electrode $i$. It is one on electrode $i$,  zero on all other electrodes, and between zero and one in the sensor volume. When a positive charge $q$ moves from position $\vec x_0$ to electrode $i$, it induces a positive current with total charge
   $Q_i = q \cdot (1 - \phi_w^i(\vec x_0))$ in electrode $i$
 and negative currents with total charges
   $Q_j = - q \cdot \phi_w^j(\vec x_0)$ in all other electrodes $j \neq i$.

 For a $p^+n$-strip sensor electrons drift to the $n^+$-rear contact and holes to the $p^+$\,strips. Electrons collected at the rear contact induce positive signals in all $p^+$\,strips. Holes however, induce a positive signal in the $p^+$\,strip on which they are collected and negative signals in all other $p^+$\,strips. If $N_q$ $eh$ pairs are generated anywhere in the sensor, and all holes are collected by strip $i$ and all electrons by the rear contact, both electron as well as hole currents induced in strip $i$ are positive, resulting in an induced charge $Q_i = N_q \cdot q_0$, where $q_0$ is the elementary charge. For strips $j$ the currents induced by the electrons are positive and the currents induced by the holes negative, resulting in a bipolar signal of total charge $Q_j = 0$.

 If not all charges are collected by the electrodes but some get stuck at position $\vec x_0$, the charge induced in a read-out strip $RO$ which does not collect holes is:

   \begin{equation}
     Q_{RO} = \int{I_{RO}} \mathrm{dt}  = (N_h\cdot q_0-N_e\cdot q_0) \cdot ( 0 - \phi_w^{RO}(\vec x_0) ) = (N_e-N_h) \cdot q_0 \cdot \phi_w^{RO}(\vec x_0),
   \end{equation}

 where $N_e$ is the number of electrons collected by the rear contact and $N_h$ the number of holes collected by the strips. For full charge collection as well as for $eh$ recombination $N_e = N_h$ and $Q_{RO} = 0$. If only electrons are lost $Q_{RO}$ is positive and negative if only holes are lost.

\begin{figure}
	\centering
  \includegraphics[width=10cm]{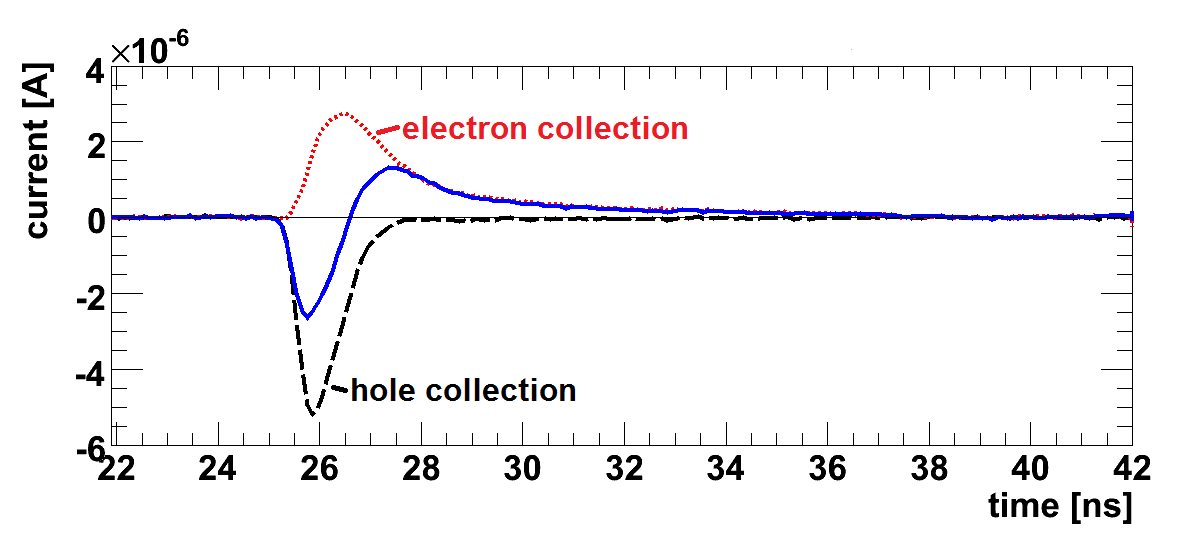}	
    \caption{Induced current in the read-out strip at $x=0$ for electron-hole pairs generated at $x=30$~$\upmu$m. Complete (solid line) and incomplete (dotted and dashed lines) charge collection are shown. For details see text.}
	  \label{fig:current}
\end{figure}

 In~\cite{Poehlsen:2012} it has been shown that the charge distribution on the sensor surface at the strip side influences the electric field in the sensor close to the Si-SiO$_2$ interface and that, depending on the biasing history and X-ray dose, situations with losses of electrons, losses of holes and no losses can be realised. This is demonstrated in Figure~\ref{fig:current} (taken from~\cite{Poehlsen:2012}) which shows three examples of currents measured on the read-out strip $RO$, positioned at $x=0$. For the coordinate system we refer to Figure~\ref{fig:sensor}. The sensor was biased to 200 V, well above the depletion voltage of 155~V, and the electron-hole pairs were generated by the laser at $x=30$~$\upmu$m just below the Si-SiO$_2$ interface.  As this position is closer to the neighbouring strip $N$ than to the readout strip $RO$, holes will only reach $RO$ by diffusion and their number will be small.

 \begin{itemize}

   \item The solid (blue) line corresponds to the situation $e = h$, where both electrons and holes are collected (non-irradiated sensor in steady state at 200~V). Holes are collected quickly ($\sim 2$~ns) at strip $N$ inducing a negative signal on the read-out strip $RO$. Electrons, which have to pass the entire sensor to reach the rear electrode, are collected more slowly ($\sim 10$~ns) inducing a positive signal in the read-out strip. Thus, the signal is bipolar and its integral is approximately zero.

   \item The dashed (black) line corresponds to the situation $e \ll h$, where mainly holes are collected (sensor irradiated to 1~MGy by $\sim 12$~keV X-rays measured after ramping the voltage from 0 to 200~V in a dry atmosphere). The positive signal of the electrons is missing and only a short negative signal induced by the holes collected at the neighbouring strip $N$ is observed.

   \item The dotted (red) line corresponds to the situation $e \gg h$, where mainly electrons are collected (non-irradiated sensor brought into a steady-state at 500~V and then biased at 200~V in a dry atmosphere). The negative signal of the holes is missing and only a positive current induced by electrons drifting to the rear side is observed.

\end{itemize}

 The following method is used for determining the charge losses:
  Light is injected sufficiently far from the read-out strip $RO$ to assure that no holes are collected by $RO$. For the measurements presented in the next chapter we choose $RO$ at $x = 0$ and inject focussed light at $x_0=75$~$\upmu$m, in the middle between two $n^+$\,strips. To check the consistency of the results data are also taken for the light injected at $x_0=40$~$\upmu$m. In addition, the signal from the rear contact is  recorded. Assuming that only one type of charge carrier is lost and $eh$ recombination can be ignored, the ``lost charge'' is
    \begin{equation}
      Q_{lost} = (N_e - N_h)\cdot q_0 = Q_{RO}/\phi_w^{RO}(\vec x_0).  \label{eq:Qlost}
    \end{equation}

  $Q_{lost}$ is positive for hole losses and negative for electron losses.   Given the quality of the bulk silicon (lifetime of the charge carriers much longer than the charge-collection time) and the fact that the sensor is operated well above the depletion voltage, charges can only be lost at the Si-SiO$_2$ interface or in a close-by accumulation layer. Thus the weighting potential at the Si-SiO$_2$ interface at the position where the $eh$ pairs are produced is used for the analysis. Compatible with the measurements presented in~\cite{Poehlsen:2012}, we use: $\phi_w^{RO}=0.05$ at $x_0=75$~$\upmu$m and 0.35 at $x_0=40$~$\upmu$m.

\section{Time dependence of charge losses}

  The measurements of the charge losses were performed for different humidities and biasing histories. For the measurements the $p^+$\,strips were grounded, either via the DC-coupled amplifiers or via 50~$\Omega$ resistors. The bias voltage was applied  at the $n^+$\,rear electrode. In~\cite{Poehlsen:2012} it has been observed for a non-irradiated sensor that, after changing the bias voltage, the distribution of surface charges is in a non-steady state, as evidenced by time dependent charge losses. It was found that the time to reach the steady state has a strong dependence on humidity. Here, we present further results on the time dependence of charge losses for the non-irradiated sensor and for the sensor irradiated to a dose of 1~MGy.

  The measurement procedure was the following: By biasing the sensor either to 0~V or to 500~V (400~V for the irradiated sensor) in a humid atmosphere ($> 60$~\% relative humidity) for at least 2 hours (typically a day and longer) steady-state conditions of the charge distribution on the surface were assumed to have been reached. Then, for the measurements under dry conditions the relative humidity was reduced. After at least 2 hours under dry conditions, the bias voltage was changed to 200~V, the charge-loss measurements started approximately 30 seconds after 200~V had been reached, and the time dependence of the lost charges determined from measurements of $Q_{RO}$ for $eh$ pairs generated by the laser at position $x_0$.

  \begin{figure}
	\centering
	 \includegraphics[width=7cm]{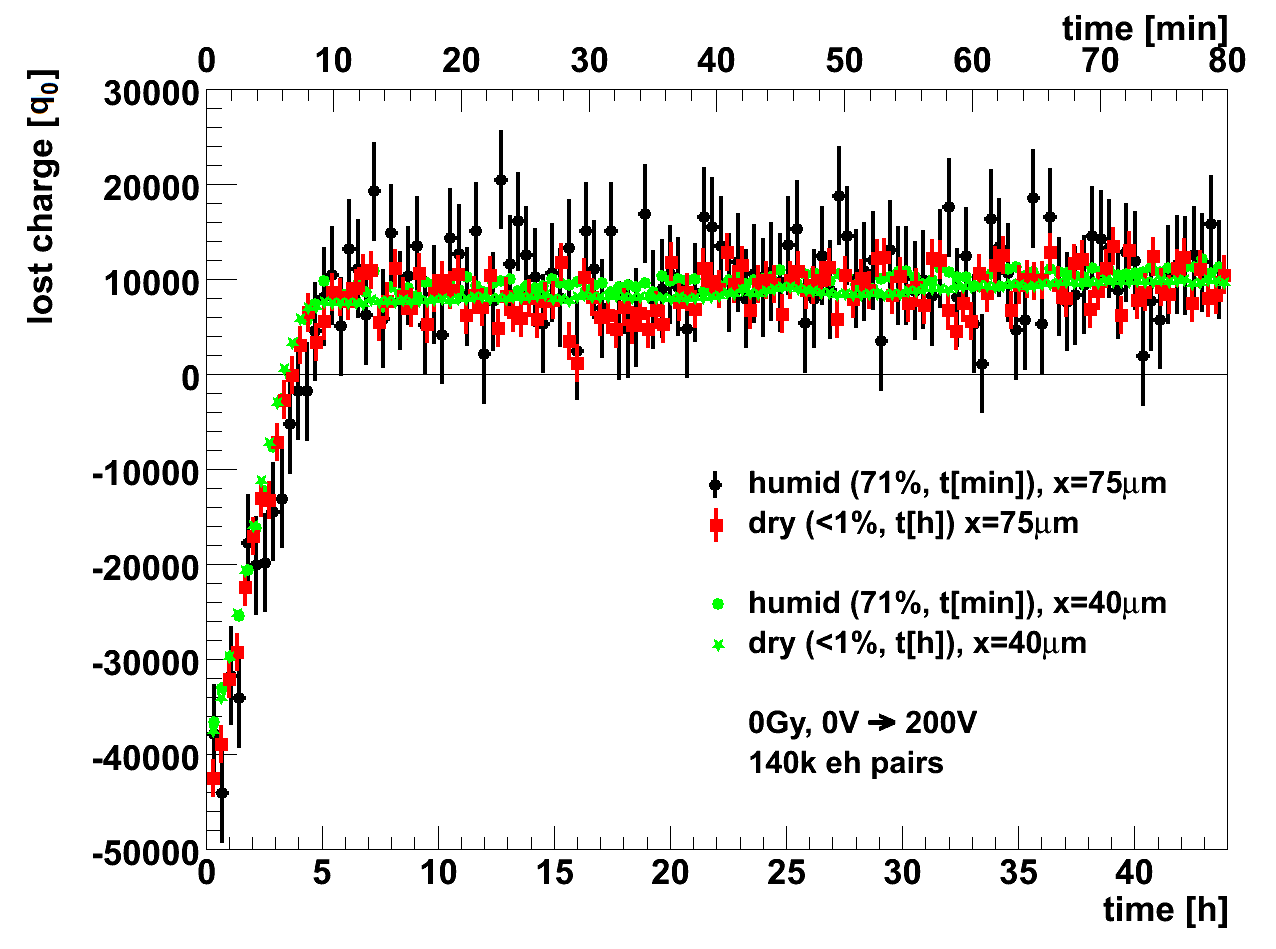}	
	 \includegraphics[width=7cm]{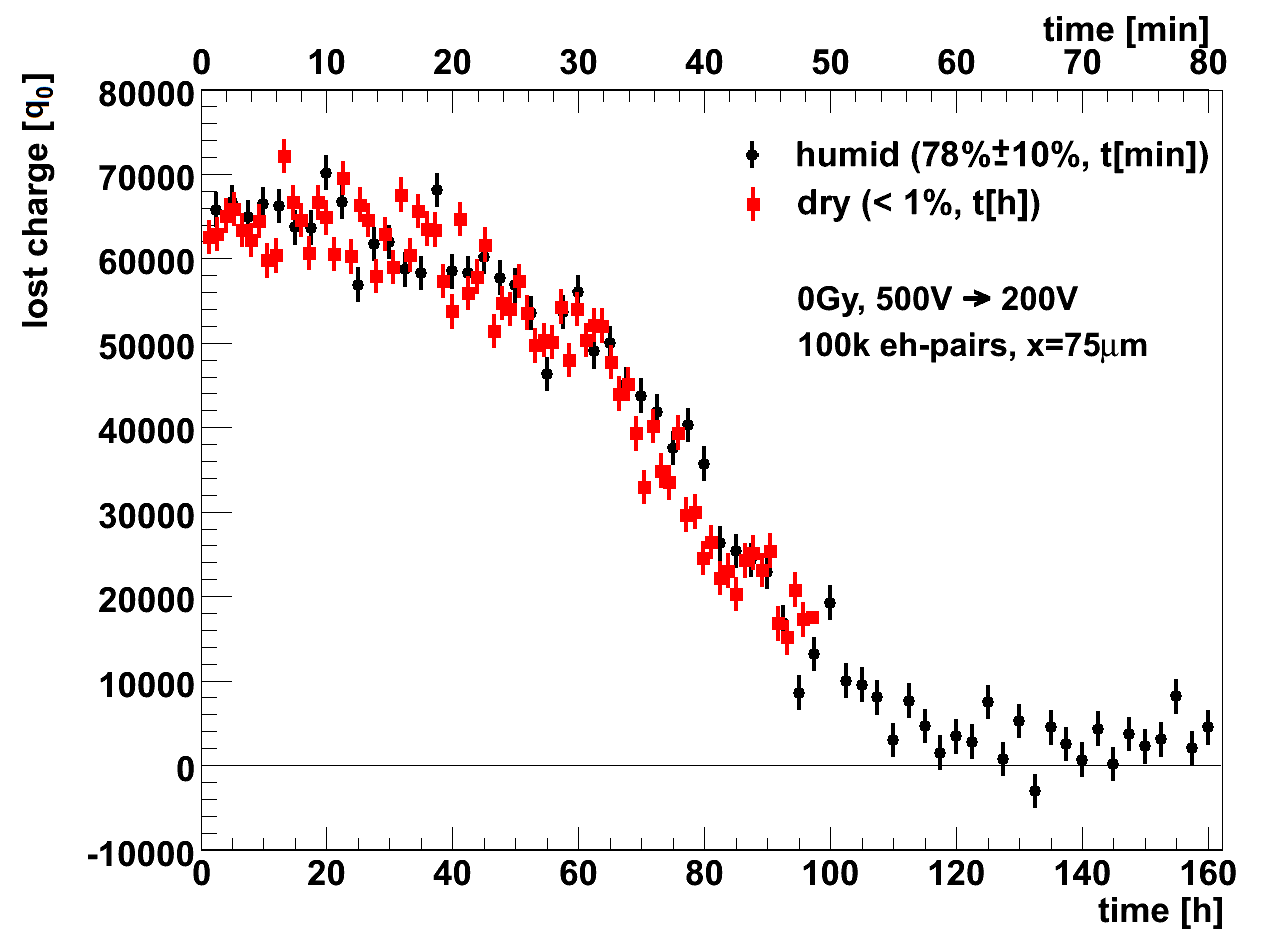}	
    \caption{Lost charge $Q_{lost}$, in units of elementary charges, as a function of time after the change of the bias voltage for the non-irradiated sensor in a dry (red rectangulars) and humid (black dots) atmosphere for light injection at $x_0=75$~$\upmu$m.
    Left: Increase of the bias voltage from 0 V to 200 V. The green curves show control measurements for light injection at $x_0=40$~$\upmu$m.
    Right: Decrease of the bias voltage from 500~V to 200~V.
    Note that the time scale for ``humid'' is on the top and for ``dry'' on the bottom.}
	\label{fig:0Gy}
\end{figure}

\begin{figure}
	\centering
	 \includegraphics[width=7cm]{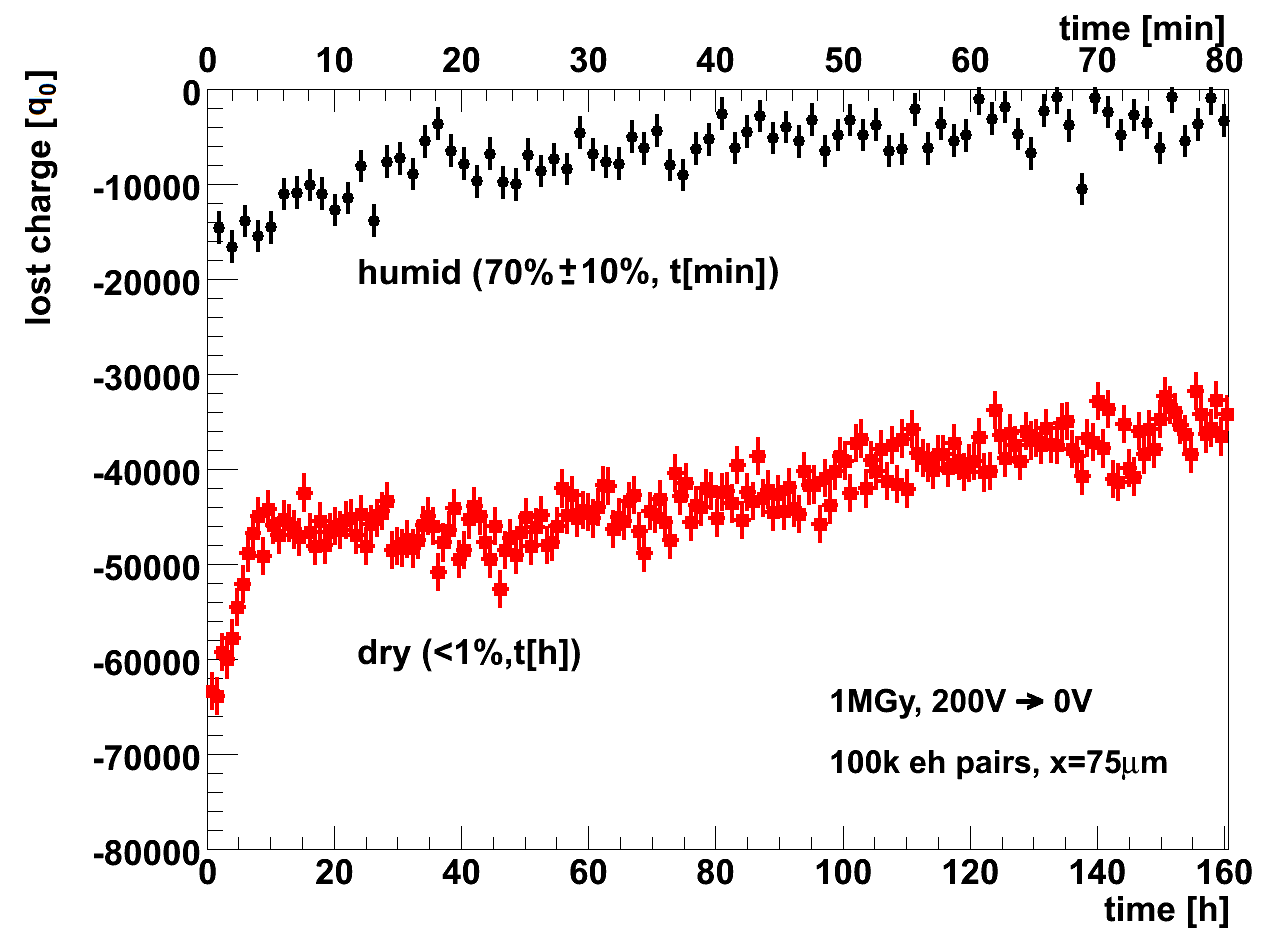}	
	 \includegraphics[width=7cm]{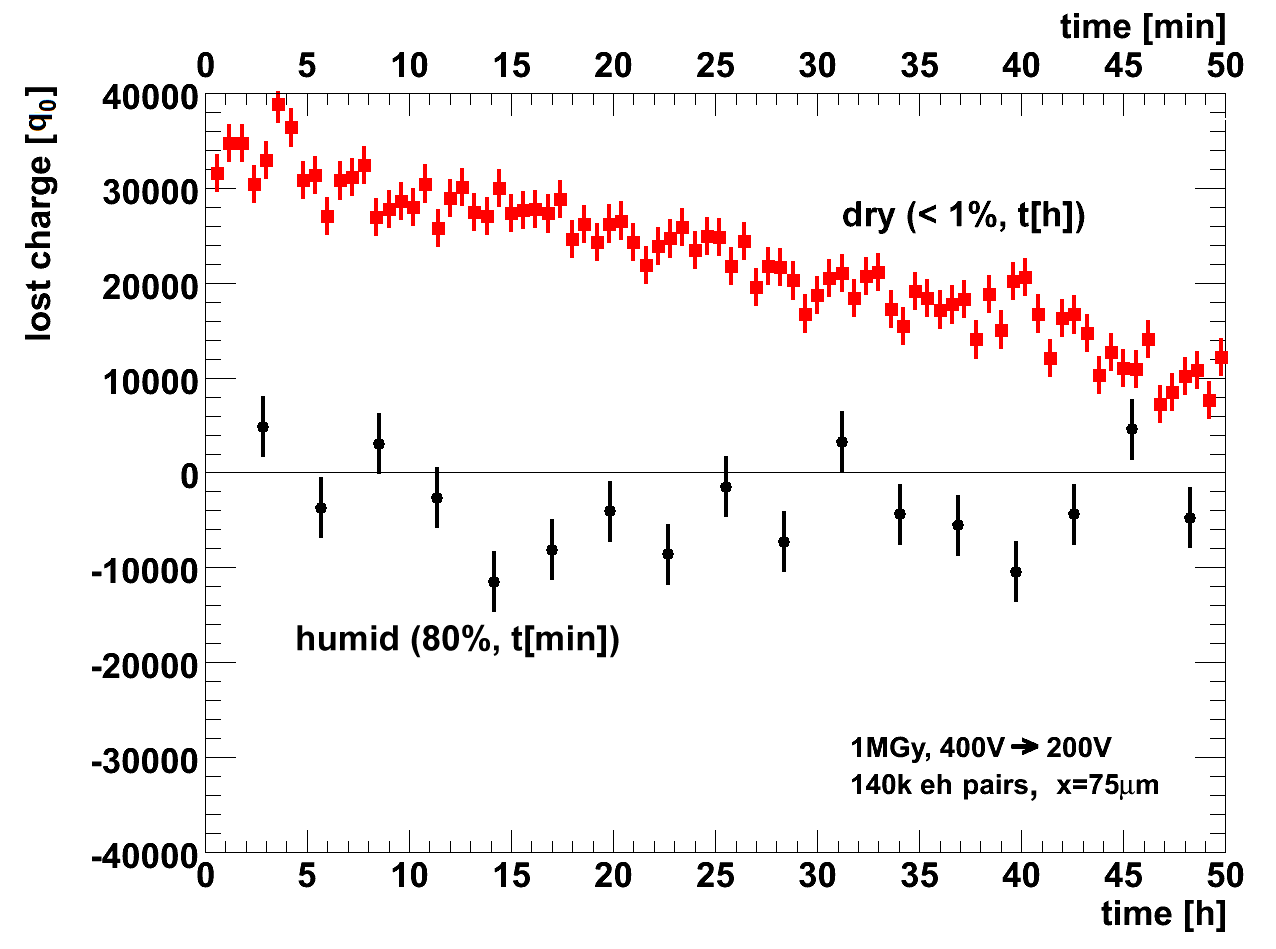}	
   \caption{Lost charge $Q_{lost}$, in units of elementary charges, as a function of time after the change of the bias voltage for the irradiated sensor in a dry (red rectangulars) and humid (black dots)  atmosphere.
   Left: Increase of the bias voltage from 0 V to 200 V.
   Right: Decrease of the bias voltage from 400~V to 200~V.
   Note that the time scale for ``humid'' is on the top and for ``dry'' on the bottom.}
	\label{fig:1MGy}
\end{figure}

  The results for the non-irradiated sensor, obtained according to Equation (\ref{eq:Qlost}), are shown in Figure~\ref{fig:0Gy}.
  When changing the bias voltage from 0 to 200~V (Figure~\ref{fig:0Gy} left) initially $\sim$~40\,000 electrons are lost for both dry and humid conditions for $\sim$~140\,000 $eh$ pairs generated by the laser. In steady-state conditions $\sim$~10\,000 holes are lost. For dry conditions (relative humidity $< 1$~\%) steady-state conditions are reached in $\sim$~6~hours, whereas for humid conditions (relative humidity $\sim 70$~\%) in $\sim$~10 minutes. For comparison, also  measurements for light injected at $x_0=40$~$\upmu$m are shown. They agree with the $x_0=75$~$\upmu$m measurements. As the weighting field at $x_0=40$~$\upmu$m is 0.35 compared to 0.05 for $x_0=75$~$\upmu$m, the  fluctuations due to noise are about an order of magnitude smaller. However, it cannot be excluded that some of the holes reach the read-out electrode $RO$ by diffusion. Therefore  we only present results for $x_0=75$~$\upmu$m in the following.

  After changing the voltage from 500~V to 200~V (Figure~\ref{fig:0Gy} right), initially $\sim$~65\,000 holes are lost for both dry and humid conditions for $\sim$~100\,000 $eh$ pairs generated. In steady-state conditions the hole losses reduce to a value between 0 and 10\,000. In humid conditions it takes $\sim 50$~minutes to reach this state, whereas in dry conditions it takes more than $\sim 100$~hours.

  We note that for dry conditions the time to reach the steady-state is about a factor 20 bigger for hole losses than for electron losses, and that the shapes of the time dependencies differ. In each case however, the shape of the time dependence for ``humid'' and ``dry'' scale. The scale factor for electron losses is $\sim 30$. For hole losses a factor $\sim 120$ is found, however, as the 78~\% humidity had an uncertainty of $\pm 10$~\% for this particular measurement, no strong conclusions can be drawn.

  Figure~\ref{fig:1MGy} shows the results for the sensor irradiated to 1~MGy, again obtained according to Equation (\ref{eq:Qlost}). After changing the voltage from 0 to 200~V electron losses are observed as for the non-irradiated sensor. The number of lost electrons however, is different for dry and humid conditions: $\sim$~65\,000 for ``dry'' and $\sim$~15\,000 for ``humid'', for $\sim$~100\,000 generated $eh$ pairs. For ``humid'' a steady state of approximately zero electron losses is reached after about 1 hour. For ``dry'' the electron losses first decrease to $\sim$~45\,000 within approximately 10 hours. Then a much slower decrease is observed, reaching $\sim$~35\,000 after 160 hours when the measurements were stopped. The current collection ring of the irradiated sensor showed a large current for voltages above 400~V. Therefore, 400~V was chosen as maximum bias voltage. The initial charge losses after reducing the voltage from 400 to 200~V are approximately zero for ``humid'', whereas 35\,000 holes are lost for ``dry'', for 140\,000 generated $eh$ pairs. As function of time the charge losses remain zero for ``humid'', whereas for ``dry'' the hole losses steadily decrease, reaching $\sim$~10\,000 after 50 hours.

  We also note that the charge losses for the irradiated sensor behave quite differently compared to the irradiated sensor. For the irradiated sensor the initial losses differ between dry and humid conditions, and the time dependencies do not scale.

\section{Discussion}

 In~\cite{Poehlsen:2012} the electron and hole losses and their time dependence have been explained with the help of detailed simulations, and their relevance for the operation of segmented $p^+n$ sensors for the detection of X-rays and charged particles discussed: Biasing a $p^+$\,sensor which has been in steady-state conditions at 0~V, results in a transverse electric field $E_y$ (see Figure~\ref{tab:sensors}) in the sensor close to the Si-SiO$_2$ interface pointing into the sensor, and a tangential field $E_x$ on the sensor surface pointing to the $p^+$\,implants. The transverse field and a possible electron accumulation layer cause the loss of electrons, and the surface field causes a movement of charges on the sensor surface until the steady-state of a uniform potential is reached. For a sensor in steady state under bias, reducing the voltage results in the opposite situation: The electric field in the sensor points towards the SiO$_2$ and therefore holes are lost. The tangential fields initially point away from the strips until by charge movement the steady-state of a uniform potential is reached. The strong dependence of the time dependencies on humidity is explained by the decrease of the surface resistivity with humidity~\cite{Grove:1967}. If the steady state is reached due to surface currents, a scaling of the time dependence with the surface conductivity is expected.

 The new measurement shown in Figure~\ref{fig:0Gy} for the non-irradiated sensor confirm the picture: Initial electron losses occur when the voltage is ramped up and hole losses when the voltage is ramped down. The initial steady-state charge losses are the same for ``dry'' and for ``humid''. The same holds for the steady-state losses. However, the time constants for reaching the steady state for hole losses is about an order of magnitude larger than for electron losses. 
This could be related to the surface charge distribution in steady state. We also note that in steady state a small hole loss of $\sim$~5~\% is observed.

 The new measurements for the sensor irradiated to an X-ray dose of 1~MGy shown in Figure~\ref{fig:1MGy} are markedly different. Again, electron losses appear when the voltage is ramped up and hole losses when it is ramped down. However, the initial losses for ``humid'' and ``dry'' are very different and the time dependencies as function of humidity do not scale. In addition, the time to reach the steady-state in dry conditions is shorter for hole losses than for electron losses. We do not have explanations for these differences, but suspect that radiation induced traps in the SiO$_2$ and the resulting high electric fields in the SiO$_2$ are the cause.

\section{Summary}

 The collection of charge carriers generated in $p^+n$-strip sensors close to the Si-SiO$_2$ interface before and after 1 MGy of X-ray irradiation has been investigated using the transient current technique with sub-nanosecond focused light pulses of 660~nm wavelength, which has an absorption length of 3.5~$\upmu$m in silicon at room temperature.  Depending on the applied bias voltage, biasing history, humidity and irradiation, incomplete collection of either electrons or holes has been observed when generating the charges at the strip side of the sensor.

 For the non-irradiated sensor electron losses are observed when the voltage is ramped up from zero to the operating voltage and hole losses when the operating voltage is reached from higher voltages. As a function of time the charge losses decrease until a steady state with hole losses between 0 and 10~\% is reached.  The shape of the time dependence and the time constant are different for hole and electron losses. At a given humidity the electron losses are about an order of magnitude faster. However, the time constants for both electron and hole losses increase by about two orders of magnitude, when the humidity is reduced from about 80~\% to below 1~\%. The observations are explained by changes of the charge distribution on the sensor surface. Initially, after changing the bias voltage the surface-charge distribution is not in a steady state, and the time constant to reach the steady state is related to the surface conductivity of positive and negative charge carriers, which are strong functions of humidity.

 For the sensor irradiated to 1~MGy the situation is more complex: As for the non-irradiated sensor, in a dry atmosphere electron losses are observed when the operating voltage is reached from lower, and hole losses when it is reached from higher voltages. The time constant for the hole losses is similar as for the non-irradiated sensor, however, the shape of the dependence is quite different. For electron losses a much longer time constant is found for the irradiated sensor. In a humid atmosphere already 30 seconds after the bias voltage is reached, the charge losses are small: approximately 15~\% electron losses when the voltage is ramped up, and less than 10~\% charge losses, when it is ramped down. We speculate that chemical effects and electric fields inside the SiO$_2$ and passivation layer due to charged traps are responsible for the differences.

 The measurements demonstrate the well-known fact that X-ray-radiation damage in the SiO$_2$ and its impact on sensor performance is highly complex.

\section*{Acknowledgements}

  This work was performed within the AGIPD Project which is partially supported by the XFEL-Company. We would like to thank the AGIPD colleagues for the excellent collaboration. Support was also provided by the Helmholtz Alliance ``Physics at the Terascale'' and the German Ministry of Science, BMBF, through the Forschungsschwerpunkt ``Particle Physics with the CMS-Experiment''. J.~Zhang is supported by the Marie Curie Initial Training Network ``MC-PAD''.


\end{document}